\pdfoutput=1
\RequirePackage{ifpdf}
\ifpdf 
\documentclass[pdftex]{sigma}
\else
\documentclass{sigma}
\fi

\newcommand{\eps}{\epsilon}

\def\cA{{\cal A}}  
\def\cD{{\cal D}}

\def\fh{{\mathfrak h}}

\newcommand{\CC}{{\mathbb C}}

\newcommand{\II}{{\mathbb I}}

\newcommand{\ZZ}{{\mathbb Z}}

\newcommand{\prt}{\partial}

\newcommand{\wt}[1]{\widetilde{#1}}

\newcommand{\half}{\frac{1}{2}}

\newcommand{\bq}{\boldsymbol{q}}

\begin{document}

\allowdisplaybreaks

\renewcommand{\PaperNumber}{079}

\FirstPageHeading

\ShortArticleName{Rational Calogero--Moser Model: Second Poisson Structure}

\ArticleName{Rational Calogero--Moser Model:
 Explicit Form\\ and $\boldsymbol{r}$-Matrix
 of the Second Poisson Structure}

\Author{Jean AVAN~$^\dag$ and Eric RAGOUCY~$^\ddag$}

\AuthorNameForHeading{J.~Avan and E.~Ragoucy}

\Address{$^\dag$~Laboratoire de Physique Th\'eorique et Mod\'elisation,\\
\hphantom{$^\ddag$}~Universit\'e de Cergy-Pontoise (CNRS UMR 8089), Saint-Martin~2,\\
\hphantom{$^\ddag$}~2, av. Adolphe Chauvin, F-95302 Cergy-Pontoise Cedex, France}
\EmailD{\href{mailto:avan@u-cergy.fr}{avan@u-cergy.fr}}

\Address{$^\ddag$~LAPTH Annecy le Vieux, CNRS and Universit{\'e} de Savoie,
\\
\hphantom{$^\ddag$}~9 chemin de Bellevue, BP 110, F-74941 Annecy-le-Vieux Cedex, France}
\EmailD{\href{mailto:ragoucy@lapp.in2p3.fr}{ragoucy@lapp.in2p3.fr}}

\ArticleDates{Received July 24, 2012, in f\/inal form October 17, 2012; Published online October 26, 2012}

\Abstract{We compute the full expression of the second Poisson bracket structure for $N=2$ and $N=3$ site
rational classical Calogero--Moser model. We propose an $r$-matrix formulation for $N=2$. It is
identif\/ied with the classical limit of the second dynamical boundary algebra previously built by
the authors.}


\Classification{81R12; 16T15; 16T25}

\section{Introduction}\label{sec:intro}

\subsection[Dynamical $r$-matrices]{Dynamical $\boldsymbol{r}$-matrices}\label{sec:dynarm}

The Calogero--Moser model~\cite{Calo,Calo+,Mo} provided~\cite{AvTa} a textbook example of a classical dynamical
$r$-matrix structure for a Lax representation~\cite{Lax} of a classical Liouville integrable system.
Remember that Liouville integrability for a $N$-dimensional Hamiltonian system is characterized (see
\cite{BaBeTa} and references therein) by the existence of $N$ independent Poisson-commuting quanti\-ties~$H^{(i)}$, $i=1, \dots, N$ including the original Hamiltonian. The system admits a Lax representation when
the equations of motion can be represented as a spectrum-preserving evolution of a $n \times n$ matrix~$L$
encapsulating the $2N$ dynamical variables $\{p_i, q_i\}$, $i=1,\dots,N$,
\begin{gather}
\frac{dL}{dt} = [L,M].
\label{Lax}
\end{gather}

The quantities $\operatorname{Tr}L^k$, $k=1,\dots,n$ may then provide the Liouville Hamiltonians
if they Poisson-commute
and they are in suf\/f\/icient number (e.g.\ if $n \geq N$ or there exists a spectral-parameter dependance).
They build in this case a so-called Hamiltonian hierarchy of mutually compatible equations of motion.
This Poisson-commuting property is equivalent~\cite{BV} to rewriting the Poisson brackets of the Lax matrix coef\/f\/icients
in a specif\/ic algebraic form involving a~mat\-rix~$r$, living in the tensor product $M_n(\CC) \otimes M_n(\CC)$,
$M_n(\CC)$ being self-explanatorily the algebra of complex $n \times n$ matrices
\begin{gather}
\{L_1, L_2\} = [r_{12}, L_1] - [r_{21}, L_2].
\label{PB1}
\end{gather}

Expression (\ref{PB1}) is the so-called linear form of $r$-matrix structure.
Associativity of the Poisson bracket form (\ref{PB1}) is guaranteed by the consistency equation
\begin{gather}
[r_{12}, r_{13}] + [r_{12}, r_{23}] +[r_{32}, r_{13}] + \{r_{12}, L_3 \} - \{r_{13}, L_2 \} =0
\label{assoc}
\end{gather}
generically known as ``dynamical Yang--Baxter equation'' (dYB), see e.g.~\cite{Mail1}. If $r$ does not
depend on the dynamical variables (\ref{assoc}) becomes a purely algebraic ``classical Yang--Baxter
equation''. If not the issue arises of an available algebraic reexpression of the dynamical contribution
$\{r_{12}, L_3 \} - \{r_{13}, L_2 \}$. At least two such forms are available in the literature. The f\/irst
one occurs in the Lax formulation of the Calogero--Moser model, where one identif\/ies
\begin{gather}
\{r_{12}, L_3 \} = \sum_{s=1}^{n} h_s^{(3)} \frac{d}{dq_s} r_{12},
\label{GNF}
\end{gather}
where $h_s$, $s=1, \dots, n$ is a representation of a Cartan subalgebra\footnote{In a more general, abstract context
the Abelian property is dropped and one is lead to consider ``non-Abelian dynamical algebras''~\cite{Ping}.}.

Note that the second one occurs in the Lax formulation of the Ruijsenaars--Schneider model~\cite{RS} and is very
much related to the f\/irst one as
\begin{gather*}
\{r_{12}, L_3 \} = \sum_{s=1}^{n} h_s^{(3)} L_3 \frac{d}{dq_s} r_{12}.
\end{gather*}

Equation (\ref{GNF}) now allows to identify the generic dYB equation (\ref{assoc}) with a specif\/ic
equation known as the classical
Gervais--Neveu--Felder equation~\cite{Feh,Fe, GeNe}. The problem of realizing an explicit algebraic form
for the dynamical terms remains however open in many interesting cases.

It is crucial to underline here that in the Lax representation (\ref{Lax}) there exists a one-to-one
algebraic correspondence, parametrized by the linear $r$ matrix,
between the Hamiltonian $H(L)$ triggering a specif\/ic time evolution, and the associated $M$ matrix. Namely
\cite{STS} the $M$ matrix is given by
\begin{gather*}
M = \operatorname{Tr}_1 \big(r_{12} dH(L)_2\big).
\end{gather*}

The Lax matrix $L$ however is characteristic of the whole Hamiltonian hierarchy and remains therefore unique.

\subsection{Hierarchy of Poisson brackets: Magri construction\label{sec:magri}}

A dual formulation of Liouville integrability was proposed by Magri~\cite{Mag} where the hierarchy of Hamiltonians
$H^{(k)} = \operatorname{Tr}L^k$, $k=1, \dots, n$ acting simultaneously on the dynamical variables~$p_i$,~$q_j$ through a single
Poisson bracket structure (e.g.\ the canonical one $\{p_i, q_j\} = \delta_{ij}$), is substituted by a hierarchy of
mutually
compatible Poisson brackets $\{ \ \}_{(i)}$, where $i=1$ corresponds to the above canonical ``f\/irst'' bracket.
The duality between the two formulations is summarized by the identity
\begin{gather*}
\big\{ H^{(k)} , X(p,q) \big\}_{(l)} = \big\{ H^{(k')} , X(p,q) \big\}_{(l')}\qquad {\rm for}\quad k+l = k'+l'.
\end{gather*}

The explicit construction of the higher ($l \geq 2)$ Poisson brackets uses the so-called ``recursion
operator'' (see, e.g.,~\cite{OeRa}).

It then follows that the Lax matrix for a given classically Liouville integrable system exhibits a~corresponding
hierarchy of $r$-matrix structures associated to each Poisson bracket in the Magri hierarchy. Such
structures are explicitly known when the $r$-matrix for the f\/irst (canonical) bracket is
non-dynamical. For instance the second Poisson bracket is the famous
quadratic Sklyanin bracket~\cite{LiPa,Mail1, Skl1}:
\begin{gather}
\{ L_1, L_2 \} = [a_{12}, L_1 L_2 ] + L_1 s_{12}L_2 - L_2 s_{12} L_1,
\label{second}
\end{gather}
where $a$ and $s$ are respectively the skew-symmetric, $ a_{12} = \half({r_{12} - r_{21}})$,
and symmetric, $s_{12} = \half({r_{12} + r_{21}})$, part of the $r$-matrix. The Sklyanin
bracket, properly said, corresponds to the case $s = 0$.

Note that a quadratic Poisson bracket (\ref{second}) takes in any case the general linear form
(\ref{PB1}) with a linear $r$ matrix def\/ined as
\begin{gather}
r_{12} \equiv \frac{1}{2} (a_{12} L_2 + L_2 a_{12}) -L_2 s_{12}.
\label{linrmat}
\end{gather}

The third bracket has a more complicated form, cubic in terms of $L$, derived explicitly in~\cite{LiPa}.

\subsection[The problem of $r$-matrix construction]{The problem of $\boldsymbol{r}$-matrix construction}

Our purpose here is to give an explicit $r$-matrix structure for the second Poisson bracket of the rational $N$-site
Calogero--Moser Lax matrix. This second Poisson structure was recently derived explicitly~\cite{BFal, Mag}
which makes it possible in principle to def\/ine an associated $r$-matrix structure. Our aim here is therefore
to provide what we believe to be the f\/irst example of $r$ matrix structure for a {\it second} Poisson
structure when the {\it first} Poisson structure is parametrized by a dynamical $r$-matrix~\cite{AvTa} in a linear
formulation~(\ref{PB1}). Indeed no explicit Sklyanin or Li--Parmentier formulation exists in such a case and
we would thus provide or disprove the existence of such a formulation in this generic case.

The Poisson bracket hierarchy however is only formulated in terms of the adjoint invariants
of the Lax matrix $I_k = \operatorname{Tr} L^k$ and mixed Lax-position matrix $J_k = \operatorname{Tr} L^k Q$ where
$Q$ is the diagonal position matrix $\operatorname{Diag} (q_1, \dots, q_N)$. Inverting this form to rexpress
the Poisson brackets of the original f\/irst-bracket canonical variables (in which the Lax matrix
elements have simple expressions) is our f\/irst aim, partially achieved in Section~\ref{section3}.

General remarks on $r$-matrix formulations are required here before we get into more details on our specif\/ic
approach. It has been established~\cite{BV} that given any f\/inite-dimensional Lax mat\-rix~$L$ together with a
Poisson structure guaranteeing Poisson commutation of the invariant traces or equivalently the eigenvalues,
one may construct a linear $r$-matrix formulation (\ref{PB1}) for the Poisson brackets of the components of $L$.
This $r$-matrix is inherently non-skew symmetric and dynamical~\cite{STS}. It is clear from example~(\ref{linrmat})
and, more subtly, from the example of the Ruijsenaars--Schneider Lax matrix~\cite{Sur}, that the linear $r$-matrix
formulation \`a la Babelon--Viallet~\cite{BV} has a priori no reason to be the most relevant or the most
adapted formulation of a~given Lax $r$-matrix structure, given in particular the algebro-geometric context
of its construction. There is actually no general rule to indicate whether a~linear, quadratic or even higher-level
$r$-matrix formulation is required, although the above mentioned algebro-geometric context may be helpful in
this respect.

Getting back to our specif\/ic issue, it appears here that although, as indicated above, one may not a priori
expect an exact Li--Parmentier procedure, we should nevertheless assume a~quadratic form such as~(\ref{second})
with arbitrary structure coef\/f\/icients $a$, $b$, $c$, $d$. In view of the formal resemblance between the second Poisson structure
on the unreduced phase space yielding the Calogero--Moser model~\cite{BFal} and the f\/irst Poisson structure on the
unreduced phase space $G \times g$ yielding the rational Ruijsenaars Schneider model (see, e.g., discussion
in~\cite{AvAnJe}), this assumption seems
all the more justif\/ied. It immediately eliminates a direct derivation \`a la Babelon--Viallet
which would only yield a cumbersome linear $r$-matrix structure from which one would have to disentangle
the $L$ dependance to get the quadratic coef\/f\/icients $a$, $b$, $c$, $d$.

Another way to derive $r$-matrix structures when the dynamical system is obtained by either a Hamiltonian
reduction (f\/irst Poisson structure of Calogero--Moser, see~\cite{OlPer}) or more generally
(our problem here, see~\cite{BFal}) a projection on reduced symplectic leaves, is to follow algebraically
the reduction of the $r$-matrix structure along the f\/ibers of the symplectic reduction. It is certainly in
principle the most signif\/icant
and illuminating way to understand the ``lower'' $r$-matrix structure once derived, and has
been achieved on a number of examples~\cite{ABT,FehPu,Pu}. It is however a long and delicate
procedure which may not be the best way for a small-sized Lax matrix. We have therefore decided to use
a direct resolution of the consistency equations obtained by inserting the explicit Poisson brackets
on the l.h.s.\ of~(\ref{second}).

\subsection{Outline of the derivation}

To obtain the Poisson brackets between physical variables $p$, $q$ we f\/irst of all (Section~\ref{section2})
derive some general Poisson bracket identities valid for all values of $N$, which allows us to reobtain
the full $N=2$ second Poisson bracket structure derived in~\cite{BFal, Mag} and construct explicitly
the $N=3$ Poisson brackets (Section~\ref{section3}). Formulae for $N=3$ are considerably more complicated and do not
suggest at this time an obvious generalization to any value of $N \geq 4$.

We then (Section~\ref{section4}) propose a completely explicit form for second Poisson bracket of the $N=2$ site
Lax matrix. This structure is def\/ined as an explicit quadratic Sklyanin form~(\ref{second})
in terms of the Lax matrix $L$. It is identif\/ied with a representation of the classical limit of the
second dynamical boundary algebra, recently built in~\cite{ARag}. It therefore does not match with the
$r$-matrix formulation of the f\/irst Poisson structure of Calogero--Moser Lax matrix,
which is in fact given by a representation of another dynamical boundary algebra,
viz. the semi-dynamical boundary algebra of~\cite{ACF1}. A direct Li--Parmentier type procedure
for dynamical $r$-matrices is therefore invalidated by this example.

The case $N=3$ seems at this time too complex and not clearly enough understood to allow for a reasonable
attempt at building an $r$-matrix structure. We conclude with some remarks and proposals.

\section[General properties of $N$-Calogero]{General properties of $\boldsymbol{N}$-Calogero}\label{section2}

The Lax matrix and position matrix for the rational $A_n$ Calogero--Moser model are def\/ined as
\begin{gather*}
L  =   \begin{pmatrix} p_1 & \frac1{q_{12}} &\ldots & \frac1{q_{1N}} \\
\frac{-1}{q_{12}} &p_2 & \ddots &\vdots \\
\vdots & \ddots & \ddots & \frac1{q_{N-1,N}} \\
\frac{-1}{q_{1N}} & \dots & \frac{-1}{q_{N-1,N}} & p_N
\end{pmatrix}\qquad\text{and}\qquad Q= \begin{pmatrix} q_1 & 0& \dots & 0 \\
0 & q_2 & \ddots & \vdots \\
\vdots & \ddots & \ddots & 0 \\
0 & \dots & 0 & q_N \end{pmatrix},
\end{gather*}
where $q_{ij}=q_i-q_j$.

The second Poisson structure is expressed~\cite{BFal} in the basis
\begin{gather*}
I_n=\frac1n \operatorname{Tr}L^n\qquad\text{and}\qquad J_{n+1}=\operatorname{Tr}\big(QL^n\big).
\end{gather*}
It reads
\begin{gather}
\{I_n,I_m\} = 0,\qquad \{J_n,I_m\} = (m+n-1)I_{m+n-1},\qquad \{J_n,J_m\} = (n-m)J_{m+n-1}.\label{eq:PB-IJ}
\end{gather}
We call this Poisson bracket algebra $\cA_N$.
Remark that the f\/irst generators $I_1$ and $J_1$ correspond to the center of mass position and momentum,
\begin{gather*}
p_0=I_1=\sum_{j=1}^N p_j\qquad\text{and}\qquad q_0=J_1=\sum_{j=1}^N q_j,
\end{gather*}
while $I_2$ is the Calogero Hamiltonian
\begin{gather*}
H= \sum_{j=1}^N p_j^2-2\sum_{j\neq i}^N\frac1{(q_{i}-q_j)^{2}}.
\end{gather*}

\subsection{Decoupling of the center of mass position}\label{sec:q0}

We extract consistently the centre-of-mass variable $q_0 \equiv J_1$ by setting
\begin{gather*}
\wt Q =Q-\frac{\operatorname{Tr}(Q)}N \II=Q-\frac{q_0}{N}\II\qquad\text{and}\qquad \wt J_{n+1}=\operatorname{Tr}(\wt QI_n)=J_{n+1} - \frac{n}{N}q_0I_n
\end{gather*}
 with the convention $\lim\limits_{n\to0} (nI_n)=N$.
\begin{proposition}
The algebra $\cA_N$ is the semi-direct sum of the subalgebra $\wt\cA_N$, generated by $I_n$ and $\wt J_n$ and the $($PB-$)$commutative algebra $\{q_0\}$
\begin{gather*}
 \{I_n,I_m\} = 0,\qquad \big\{\wt J_{n+1},I_m\big\} = (m+n)I_{m+n}-\frac{mn}{N} I_mI_n,\\
  \big\{\wt J_{n+1},\wt J_{m+1}\big\} = (n-m)\wt J_{m+n+1} +
\frac{mn}{N}\big( \wt J_{m+1}I_n - \wt J_{n+1}I_m\big),\\
 \{q_0,I_m\} = m I_m,\qquad \big\{q_0,\wt J_{n+1}\big\} = n\wt J_{n+1}.
\end{gather*}

The structure of the algebra $\wt\cA_N$
is entirely determined by the PBs of the $p_i$'s and the $q_{ij}$'s and does not depend on the PBs of $q_0$.
\end{proposition}

\begin{proof}
The PBs are obtained by direct calculation. It follows that, since $\wt Q$ and $L$ do not depend on $q_0$,
the PBs of $q_0$ with $I$ and $J$ variables are not relevant in the calculation of the $L$ Poisson brackets
and the associated $r$-matrix structure. They shall be considered separately.
\end{proof}

\subsection{PBs of the center of mass momentum}\label{sec:p0}
\begin{proposition}
The Poisson brackets of $p_0$ are given by
\begin{gather}
\{p_0,q_j\} = -p_j\qquad\text{and}\qquad \{p_0 , p_j\} = -2 \sum_{n\neq j} q_{jn}^{-3}.
\label{eq:PBp0}
\end{gather}
\end{proposition}
\begin{proof}
Let
\begin{gather*}
 K= \begin{pmatrix} a_1 & q_{12}^{-2} &\ldots & q_{1N}^{-2} \\
q_{12}^{-2} &a_2 & \ddots &\vdots \\
\vdots & \ddots & \ddots & q_{N-1,N}^{-2} \\
q_{1N}^{-2} & \dots & q_{N-1,N}^{-2} & a_N
\end{pmatrix} \qquad\text{with}\quad
a_j= -\sum_{n\neq j} q_{nj}^{-2},
\end{gather*}
then (\ref{eq:PBp0}) is equivalent to
\begin{equation*}
\{p_0, L\} = [L,K]\qquad\text{and}\qquad \{p_0,Q\} = [Q,K]-L.
\end{equation*}

This matricial form of the PBs implies
\begin{equation*}
\{p_0, I_n\} = \sum_{m=0}^n \operatorname{Tr} L^m [L,K] L^{n-m-1} = \operatorname{Tr}\big(L^{n-1}K-KL^{n-1}\big)= 0
\end{equation*}
and
\begin{gather*}
\{p_0, J_{n+1}\}  =  \operatorname{Tr}\big( [Q,K]-L\big)L^n+
\sum_{m=0}^{n+1} \operatorname{Tr} Q L^m [L,K] L^{n-m} \\
\hphantom{\{p_0, J_{n+1}\}}{}
 =  \operatorname{Tr}\big(KQL^{n}-QL^{n}K\big) -\operatorname{Tr}L^{n+1} = -(n+1) I_{n+1}.
\end{gather*}
Hence, (\ref{eq:PBp0}) reproduce the PBs of $p_0$ with all the generators of~$\cA_N$. Since the correspondence
between $I$, $J$ and $p$, $q$ variables is one-to-one~\cite{BFal}, the PB's of $p_0$
with $q_i$, $p_j$ are univocally determined by the PB's of~$p_0$
with $I_i$, $J_j$ establishing that~(\ref{eq:PBp0}) is correct.
\end{proof}
It is interesting to remark that this matrix $K$ (used above) is such that the sum of elements on every line
or every column yields $0$. It implies that $K$ commutes with the matrix $\mu \equiv \sum\limits_{i \neq j} e_{ij}$
which is the moment map used to def\/ine the Calogero--Moser model by Hamiltonian reduction of a~free motion
on the cotangent bundle of the Lie algebra $M_n(\CC)$~\cite{OlPer}. However the corresponding Poisson
structure is the {\it first}, not {\it second} one and the meaning of this property of $K$ is therefore not
clear.

\begin{corollary}\label{corollary1}
For any function of $\vec q$ and $\vec p$, we have
\begin{equation*}
\{p_0,f(\vec q,\vec p\,)\}=\cD f(\vec q,\vec p\,) \qquad
\text{with} \quad \cD=\sum_{n=1}^N\left\{ p_n\frac\prt{\prt q_n} -2\sum_{j\neq n} q_{jn}^{-3}\frac\prt{\prt q_j}
\right\}.
\end{equation*}
\end{corollary}

\begin{proof}
Direct calculation using (\ref{eq:PBp0}). \end{proof}

\section[Second Calogero-Poisson brackets]{Second Calogero--Poisson brackets}\label{section3} 

\subsection[$N=2$ Poisson brackets]{$\boldsymbol{N=2}$ Poisson brackets}\label{calo2}

These second PBs have been already calculated in~\cite{BFal}. We recall them for
the sake of completeness and show on a simple case the method we use for $N=3$.

From the calculation of Section~\ref{sec:p0}, one deduces that
\begin{equation*}
\{p_1,p_2\}=-\frac1{q_{12}^3}
\qquad\text{and}\qquad \{p_j,q_k\}=-\delta_{jk}p_j+z_{jk}
\qquad \text{with} \quad z_{1j}+z_{2j}=0\quad \forall\, j,
\end{equation*}
which leaves us with three unknowns, $z_{11}$, $z_{22}$ and the PB $\{q_1,q_2\}$.

Plugging this partial result into the PBs given in (\ref{eq:PB-IJ}),
one deduce the f\/inal form, parametrized as
\begin{gather}
 z_{11} = -z_{22}= -z_{21} = z_{12} = \frac{(p_1 - p_2)}{(q_1 - q_2)^2}
\frac{1}{\frac{4}{(q_1 - q_2)^2}- (p_1 -p_2)^2}, \nonumber\\
 \{q_1,q_2\}= \frac{1}{(q_1 - q_2)} \frac{1}{\frac{4}{(q_1 - q_2)^2}-(p_1 -p_2)^2}.
\label{PB2site}
\end{gather}

As already mentioned, these expressions were calculated
in~\cite{BFal}, directly from the PBs of the $I_n$'s and $J_m$'s. However,
a direct calculation becomes highly complicated for larger values of $N$, and one needs to use
the results obtained in Sections~\ref{sec:q0} and~\ref{sec:p0}. This def\/ines the strategy we
will adopt in the next subsection.

\subsection[$N=3$ Calogero]{$\boldsymbol{N=3}$ Calogero}\label{sec:calo3}

We set the following forms for the $\{p,p\}$ and $\{p,q\}$ brackets
\begin{gather*}
 \{p_i,p_j\}=-\frac{2}{q_{ij}^3} +x_{ij}\qquad\text{and}\qquad \{p_i,q_{jk}\}=(\delta_{ik}-\delta_{ji})p_i +z_{i;jk}, \\
 \text{with} \quad x_{ij}=-x_{ji}\qquad\text{and}\qquad z_{i;jk} =-z_{i;kj}.
\end{gather*}

A careful study of Jacobi identity for the triplet $p_0$, $p_1$, $p_2$ f\/inally yields
\begin{gather*}
x_{12}=x_{23}=x_{31}\equiv x_0.
\end{gather*}

Consistency conditions on the $z$ coef\/f\/icients read
\begin{gather*}
 z_{1;jk} +z_{2;jk} +z_{3;jk} =0\quad \forall\, j\neq k,\qquad  z_{i;12} +z_{i;23} +z_{i;31} =0\quad \forall\, i.
\end{gather*}

Explicit resolution of the Poisson bracket structure can then be achieved and yields
\begin{gather*}
z_{i;jk}  =  -\left(\frac{q_{jk}}{q_{ii_2}q_{ii_3}}\right)^3 \frac{n_{i;jk} }{\mathsf{d}}
\qquad \text{with} \quad (i,i_2,i_3)=\text{circ.perm.}(1,2,3).
\end{gather*}
Two forms of $n$ coef\/f\/icients are def\/ined depending on which independent indices are present. One gets
\begin{gather*}
n_{i;jk}  =  -q_{ij} q_{ik}\left\{ q_{ij}^2 q_{ik}^2 \big({-}q_{ij}^3 p^2_j +q_{ik}^3p_k^2\big)
+p_i (p_j -p_k) q_{ij}^2 q_{ik}^2 \left(q_{ij} + \half q_{ik}\right) {\bq}^2 \right.\\
\left. \hphantom{n_{i;jk}  =}{}
 - p_jp_k q_{ij}^2 q_{ik}^2 q_{jk} \big(q_{jk}^2 + 3q_{ik}q_{ij}\big) + \half{\bq}^2 q_{jk} \big(q_{ij} +q_{ik}\big)^2\right\} x_0  - {\bq}^2 (q_{ij} +q_{ik})^2,
\\
n_{i;ij}  =  \left\{-(q_{ij}q_{jk}q_{ki})^3(p_j -p_i)(p_j -p_k) + \half {\bq}^2 q_{ij}q_{ik}q_{jk}(q_{ij} +
q_{ik})(q_{ik}+q_{jk})\right\}x_0\\
 \hphantom{n_{i;ij}  =}{} +4q_{ik}^3-2q_{ij}q_{jk}\big(q_{ik}^2 + q_{ij}q_{jk}\big).
\end{gather*}
Here it is understood that $(i,j,k)$ is any permutation of the three indices $1$, $2$, $3$. In addition
one def\/ines
\begin{gather*}
\mathsf{d}  =  -\bq^2\big(p_1(q_{12}+q_{13})+p_2(q_{23}+q_{21})+p_3(q_{31}+q_{32})\big),
\qquad
\bq^2  =  q_{12}^2+q_{23}^2+q_{13}^2.
\end{gather*}
Finally the single, pure $q_{ij}$ bracket reads
\begin{gather*}
\{ q_{12}, q_{23} \}  =  - \frac{n_{12}n_{23}}{2\mathsf{d}},\\
n_{12}n_{23}  =  q_{12}q_{13}q_{23} x_0 \big\{ (q_{12}q_{13}q_{23})^2(p_1p_2(p_1-p_2) + p_2p_3(p_2-p_3) + p_1p_3(p_1-p_3))
\\
\hphantom{n_{12}n_{23}  =}{}
 +q_{23}^2 (q_{12}-q_{31})p_1 - q_{13}^2 (q_{23}-q_{12})p_2 - q_{12}^2 (q_{13}+q_{23})p_3
\big\}
\\
\hphantom{n_{12}n_{23}  =}{}
  +{\bq}^2 \big\{p_1q_{23}(q_{12} + q_{13}) + p_2q_{13}(q_{12} + q_{32})
 - p_3q_{12}(q_{13} + q_{23}) \big\}.
\end{gather*}

The quantity $x_0$ is explicitly obtained as a very complicated rational function of all
dynamical variables $p$ and $q_i -q_j$.
It can be characterized however as the single solution
of the dif\/ferential equation ($\cD$ being def\/ined in Corollary~\ref{corollary1})
\begin{gather*}
\cD x_0  =  \frac{x_0}{q_{12}q_{23}q_{31}\mathsf{d}}\bigg\{
 q_{23}\big(q_{12}^4+q_{13}^4\big)p_1^2
+q_{31}\big(q_{21}^4+q_{23}^4\big)p_2^2 +q_{12}\big(q_{31}^4+q_{31}^4\big)p_3^2
\\
\hphantom{\cD x_0  =}{}
+p_2p_3q_{23}\big(2q_{23}^4+3q_{12}q_{23}^2q_{13}-q_{12}^2q_{13}^2\big)
+p_1p_2q_{12}\big(2q_{12}^4+3q_{23}q_{12}^2q_{13}-q_{23}^2q_{13}^2\big)
\\
\hphantom{\cD x_0  =}{}
+p_1p_3q_{31}\big(2q_{13}^4-3q_{23}q_{13}^2q_{12}-q_{23}^2q_{12}^2\big)
+\frac{(\bq^2)^3}{4q_{12}q_{23}q_{31}} \bigg\}
\\
\hphantom{\cD x_0  =}{}
+ \frac{2\big(q_{13}^6 +q_{12}^6+q_{23}^6\big) -6q_{23}^2q_{12}^2q_{13}^2}{(q_{12}q_{23}q_{31})^3\mathsf{d}},
\end{gather*}
with the particular value at $q_{12}=q_{23}$
\[
x_0\big|_{q_{12}=q_{23}}  =  -18 \frac{ 81-3q_{12}^2(p_3+p_1-2p_2)^2-4q_{12}^4(p_2-p_1)(p_1-p_3)^2(p_3-p_2)}
{q_{13}q_{12}^2g(q_{12})g(-q_{12})}
\]
with
\begin{gather*}
g(q_{12})= 27+27q_{12}(p_3-p_1)+4q_{12}^2(2p_3-p_2-p_1)(p_2+p_3-2p_1)\\
\phantom{g(q_{12})=}
{}+4q_{12}^3(p_2-p_1)(p_3-p_1)(p_3-p_2).
\end{gather*}
Note also that
\begin{equation*}
x_0 \sim \frac{-2}{q_{23}q_{13}q_{12}}\qquad\text{for}\quad q_{12}\to 0
\end{equation*}
for $q_{23}$ f\/inite.

\subsection{Particular cases}

We present here some particular cases where the PBs simplify drastically. They correspond to
particular positions of the particles. Remark that in some cases, this choice of particular positions make
the Lax formalism ill-def\/ined, but the PBs are themselves well-def\/ined.

\subsubsection{Three free particles}
If we consider $q_{12}=q_{23}\to\infty$, the three particles are far away one from each other,
so that they can be considered as decoupled. Indeed, in that case, the PBs simplify to
\begin{gather*}
\{q_i,q_j\}=0,\qquad \{p_i,q_j\}=-\delta_{ij}p_i,\qquad \{p_i,p_j\}=-\frac{2}{q_{ij}^3}.
\end{gather*}
One recognizes the second PB structure of free particles.

Let us note that in order to take properly the limit, one has to explicitly use the behavior
\begin{equation*}
x_0\sim \frac9{4(p_2-p_1)(p_2-p_3)q_{12}^5}\qquad\text{when}\quad  q_{23}=q_{12}\to\infty.
\end{equation*}

\subsubsection{One free particle}
If we now consider that only one particle, say particle 1, is far from the two others, one
has to take the limit $q_{12}\to\infty$ keeping $q_{23}$ f\/inite. The f\/irst particle
(associated to the index 1) decouples while the particles 2 and 3 still interact. Indeed, in this case,
the PBs simplify to
\begin{gather*}
 \{q_1,q_j\}=0,\qquad \{p_j,q_1\}=-\delta_{1j}p_1,\qquad \{p_1,q_j\}=-\delta_{1j}p_1,\qquad \{p_1,p_j\}=-\frac{2}{q_{1j}^3}\quad\forall\, j,
\\
 \{q_2,q_3\}= \frac{-2\,q_{23}}{q_{23}^2(p_2-p_3)^2-4},\qquad \{p_2,p_3\}=-\frac{2}{q_{23}^3},
\\
 \{p_j,q_k\}=-\delta_{kj}p_j+ z_{kj},\qquad j,k=2,3,
\\
 z_{2;12}=-\frac{p_2-p_3}{q_{23}^2(p_2-p_3)^2-4},
  \qquad
  z_{1;12}=0,\qquad z_{1;23}=0,\\
\{q_{12},q_{23}\}= -\frac{2q_{23}}{q_{23}^2(p_2-p_3)^2-4}.
\end{gather*}

One recovers indeed $z_{jk}$ such as computed in (\ref{PB2site}).
Again, one needs to know that for $q_{12}\to\infty$,
\begin{gather*}
x_0 \sim \frac{2}{q_{23}^2(p_{1}^2-p_{1}(p_{3}-p_{2})+p_{2}p_{3})+1}\!\left(\!
\frac{-1/q_{23}}{(q_{23}+q_{12})q_{12}}
+\frac{(p_{3}-p_{2})(2p_{1}-p_{3}-p_{2})q_{23}^2}{(q_{23}^2(p_{3}-p_{2})^2-4)
(q_{23}+q_{12})q_{12}^2}\!\right)
\end{gather*}
to get a correct answer.

\section[Dynamical $r$-matrix algebra for $N=2$ Calogero]{Dynamical $\boldsymbol{r}$-matrix algebra for $\boldsymbol{N=2}$ Calogero}\label{section4} 

\subsection{The quadratic algebra}\label{sec:quadalg}

Let us now formulate the Poisson bracket structure~(\ref{PB2site}) in terms of an $r$-matrix structure.
We postulate that a quadratic formulation mimicking~(\ref{second}) will be adequate for this second
Poisson bracket, although both~$a$ and~$s$ matrices will be expected to be dynamical.
We recall that given a classical Lax matrix~$\ell$, the most general quadratic form for the
associated Poisson structure is
\begin{gather}
\{ \ell_{1}, \ell_{2}\} = a_{12} \ell_{1}\ell_{2}+
\ell_{1}b_{12}\ell_{2}-\ell_{2}c_{12}\ell_{1}
-\ell_{1}\ell_{2}d_{12},
\label{eq:dyn-class}
\end{gather}
where consistency conditions imply that $a_{12}=-a_{21}$,
$d_{12}=-d_{21}$, $b_{12}=c_{21}$.
Note that (\ref{eq:dyn-class}) implies that the functions
$\{\operatorname{tr}\ell^m,\; m\in\ZZ_{+}\}$ Poisson-commute if $a+b=c+d$. A
more general trace formula,
$\operatorname{tr}(\gamma^{-1}\ell)^m$, occurs whenever a scalar matrix~$\gamma$
exists such that
\begin{gather*}
a_{12} \gamma_{1}\gamma_{2}+
\gamma_{1}b_{12}\gamma_{2}-\gamma_{2}c_{12}\gamma_{1}
-\gamma_{1}\gamma_{2}d_{12}
=0,
\end{gather*}
see~\cite{MaFrei1}.

Dynamical dependence of $abcd$ now is assumed to be solely on coordinates
$q_{i}$, $i=1,\ldots,n$, on a dual $\fh^*$ of the Cartan subalgebra
$\fh$ in $sl(n,\CC)$.

In the 2 sites case we get
\begin{gather*}
\ell =  \begin{pmatrix}
 {p_1}&  \dfrac{1}{q_{12}} \vspace{1mm}\\
\dfrac{-1}{q_{12}} &  {p_2}
\end{pmatrix} .
\end{gather*}
The PB deduced from (\ref{PB2site}) reads
\begin{gather}
\{ \ell_{1}, \ell_{2}\} = \frac{1}{q_{12}}  \begin{pmatrix}
0 & \dfrac{p_1}{q_{12}}& -\dfrac{p_1}{q_{12}} &0\\[2ex]
- \dfrac{p_1}{q_{12}} &  -\dfrac{2}{(q_{12})^2} & 0 & \dfrac{p_2}{q_{12}}\\[2ex]
 \dfrac{p_1}{q_{12}} & 0 & \dfrac{2}{(q_{12})^2} &  -\dfrac{p_2}{q_{12}}\\[2ex]
0 & - \dfrac{p_2}{q_{12}}&\dfrac{p_2}{q_{12}} &0
\end{pmatrix} .
\label{eq:pbL}
\end{gather}

As in the linear case (\ref{assoc}), (\ref{GNF}), associativity for the PB structure
(\ref{eq:dyn-class}) is implied by
algebraic consistency conditions (Yang--Baxter classical equations)
for $a$, $b$, $c$, $d$, provided the \textit{a priori} undetermined bracket
$\{r_{12},\ell_{3}\}$, $r=a,b,c,d$, be of an algebraic form. We
postulate here the following form for this PB
\begin{gather}
\{r_{12},\ell_{3}\}  =  \eps_{R}(h_{3} \prt r_{12})\ell_{3}+
\eps_{L}\ell_{3}h_{3} \prt r_{12},
\label{eq:rl-class}\\
h\prt  =  \sum_{i=1}^n \mu e_{ii}\otimes \frac\prt{\prt q_{i}},
\label{eq:h-prt}
\end{gather}
where $e_{ii}\in\fh$, $\eps_{R}$, $\eps_{L}$ are $c$-numbers to be determined. Notice the
dif\/ference in homogeneity factors in $l$ with respect to the linear case~(\ref{GNF}).

We will see below that this postulate is consistent and the correct choice of parameters~$\eps$ is
\begin{gather*}
\eps_L=\eps_R=\half.
\end{gather*}
A solution to express the PB (\ref{eq:pbL}) as a quadratic form is given by
\begin{gather*}
a_{12}  =  d_{12}=  \begin{pmatrix}
0 & 0& 0 &0\\
0 &\dfrac{-w_1}{2q_{12}} & \dfrac{1}{2q_{12}} & 0 \vspace{1mm}\\
0 & \dfrac{-1}{2q_{12}} & \dfrac{w_1}{2q_{12}} & 0 \\
0 & 0 & 0 &0
\end{pmatrix} ,\qquad
b_{12}  =  c_{21}=c_{12}= \begin{pmatrix}
  \dfrac{-w_1}{2q_{12}} &0& 0 &  \dfrac{1}{2q_{12}} \\
0 & 0& 0 &0\\
0 & 0 & 0 &0\\
\dfrac{-1}{2q_{12}} &0 & 0 & \dfrac{w_1}{2q_{12}}
\end{pmatrix} ,
\end{gather*}
where $w_1$ is a free parameter.

Jacobi identity for the quadratic PB is then ensured, given (\ref{eq:rl-class})
by the following classical dynamical Yang--Baxter equations
\begin{gather}
{[a_{12},a_{13}]}+{[a_{12},a_{23}]}+{[a_{32},a_{13}]}+
\half\big( h_{3}\prt a_{12}+h_{1}\prt a_{23}+h_{2}\prt a_{31}\big)
=0 , \label{eq:class-dYBE-a}\\
{[d_{12},d_{13}]}+{[d_{12},d_{23}]}+{[d_{32},d_{13}]}+
\half\big( h_{3}\prt d_{12}+h_{1}\prt d_{23}+h_{2}\prt d_{31}\big)
=0 , \label{eq:class-dYBE-b}\\
{[a_{12},c_{13}+c_{23}]}+{[c_{13},c_{23}]}+
\half\big({-}h_{3}\prt a_{12}+h_{1}\prt c_{23}-h_{2}\prt
c_{13}\big)
=0 ,\label{eq:class-dYBE-c} \\
{[d_{12},b_{13}+b_{23}]}+{[b_{13},b_{23}]}+
\half\big({-}h_{3}\prt d_{12}+h_{1}\prt b_{23}-h_{2}\prt
b_{13}\big)
=0.\label{eq:class-dYBE-d}
\end{gather}
In the absence of dynamical term, one would recover the usual classical
quadratic algebra~\cite{MaFrei1}.

A connection between the classical DYB equations
associated with quadratic and linear Poisson brackets is established as follows: In the
simplest case of Poisson commutation of traces when $a+b = c+d$ in~(\ref{eq:dyn-class})
the matrix $r \equiv a+b$ obeys the linear dynamical Yang--Baxter equation obtained from
(\ref{assoc}), (\ref{GNF}), that is, with two derivative terms provided that $a$, $b$, $c$, $d$ obey
(\ref{eq:class-dYBE-a})--(\ref{eq:class-dYBE-d}), coupled DYB with three derivative terms.

This result however does {\it not} imply that (\ref{assoc}) and~(\ref{eq:dyn-class})
are identif\/ied as consistent f\/irst and second Poisson structure for $l$ since the algebraic forms
(\ref{eq:class-dYBE-a})--(\ref{eq:class-dYBE-d})
require in addition respectively the identif\/ication of the Poisson
brackets of $l$ with $r$ or $a$, $b$, $c$, $d$ as~(\ref{GNF}) or~(\ref{eq:rl-class}) which is not
implied in any way by the form of PB's (\ref{assoc}) and (\ref{eq:dyn-class}).
In other words, contrary to the non-dynamical case (Sklyanin bracket) one cannot establish that
the quadratic form (\ref{eq:dyn-class}) be a consistent Poisson structure
{\it solely} from the fact that (\ref{assoc}) be one such structure even if
(\ref{eq:class-dYBE-a})--(\ref{eq:class-dYBE-d}) hold.
One then immediately observes that
(\ref{eq:class-dYBE-a})--(\ref{eq:class-dYBE-d}) is a classical
limit ($\hbar\to0$) of a set of 4 dynamical Yang--Baxter equations f\/irst formulated in~\cite{ARag}
\begin{gather}
A_{12}(\boldsymbol q)A_{13}\big(\boldsymbol q-\eps_{R}h^{(2)}\big)A_{23}(\boldsymbol q)
 = A_{23}\big(\boldsymbol q-\eps_{R}h^{(1)}\big)A_{13}(\boldsymbol q)
A_{12}\big(\boldsymbol q-\eps_{R}h^{(3)}\big),\label{eq:dynYBE-a}\\
D_{12}\big(\boldsymbol q+\eps_{L}h^{(3)}\big)D_{13}(\boldsymbol q)
D_{23}\big(\boldsymbol q+\eps_{L}h^{(1)}\big)
 = D_{23}(\boldsymbol q)D_{13}\big(\boldsymbol q+\eps_{L}h^{(2)}\big)
D_{12}(\boldsymbol q),
\label{eq:dynYBE-b}\\
A_{12}(\boldsymbol q)C_{13}\big(\boldsymbol q-\eps_{R}h^{(2)}\big)C_{23}(\boldsymbol q)
 = C_{23}\big(\boldsymbol q-\eps_{R}h^{(1)}\big)C_{13}(\boldsymbol q)
A_{12}\big(\boldsymbol q+\eps_{L}h^{(3)}\big),
\label{eq:dynYBE-c}\\
D_{12}\big(\boldsymbol q-\eps_{R}h^{(3)}\big)B_{13}(\boldsymbol q)
B_{23}\big(\boldsymbol q+\eps_{L}
h^{(1)}\big)
 = B_{23}(\boldsymbol q)B_{13}\big(\boldsymbol q+\eps_{L}h^{(2)}\big)
D_{12}(\boldsymbol q),
\label{eq:dynYBE-d}
\end{gather}
with the particular choice of ``weight parameters'' $\eps_L=\eps_R=\half$. The classical limit
is def\/ined by setting
\begin{gather*}
R(\boldsymbol q)  =  \II +\hbar r(\boldsymbol q)+o\big(\hbar^2\big),\qquad
R=A,B,C,D\qquad\text{and}\qquad r=a,b,c,d,
\\
h^{(i)}  =  \hbar h_{i}+o\big(\hbar^3\big),
\end{gather*}
and keeping the order $\hbar^2$ in
(\ref{eq:dynYBE-a})--(\ref{eq:dynYBE-d}), orders $1$ and $\hbar$
being trivial.

These 4 equations are in turn characterized as suf\/f\/icient conditions
for associativity of a~quantum quadratic dynamical exchange algebra def\/ined generically in~\cite{ARag}
\begin{gather*}
A_{12}(\boldsymbol q)K_{1}\big(\boldsymbol q-\eps_{R}h^{(2)}\big)B_{12}(\boldsymbol q)
K_{2}\big(\boldsymbol q+\eps_{L}h^{(1)}\big)\\
\qquad{}
 = K_{2}\big(\boldsymbol q-\eps_{R}h^{(1)}\big)C_{12}(\boldsymbol q)
K_{1}\big(\boldsymbol q+\eps_{L}h^{(2)}\big)D_{12}(\boldsymbol q)
\end{gather*}
assuming a set of zero-weight conditions
\begin{gather*}
 \eps_{R}\big[h^{(1)}+h^{(2)},A_{12}\big] =\eps_{L}\big[h^{(1)}+h^{(2)},D_{12}\big] =0,
\\
 \big[\eps_{R}h^{(1)}-\eps_{L}h^{(2)},C_{12}\big] = \big[\eps_{L}h^{(1)}-\eps_{R}h^{(2)},B_{12}\big] =0,
\end{gather*}
and unitary hypothesis
\[
  A_{12}A_{21}=D_{12}D_{21}=\II\otimes\II, \qquad C_{12}=B_{21}.
\]
Altogether, these relations ensure associativity of the product in
the dynamical algebra.

Note that the free parameter $w_1$ is the signature in this classical limit of one particular gauge covariance
of the dynamical Yang--Baxter equation for $A$, pointed out, e.g., in~\cite{ABilRol} under which diagonal
coordinates $d_{ij}$ of $R$-matrix on basis elements $e_{ij} \otimes e_{ji}$ in $M_n(\CC) \otimes M_n(\CC)$ contain
constant parameters $g_{ij} \equiv f_i - f_j$ and $f_i$ are arbitrary non-dynamical c-numbers. Here
$ \hbar w_1 \equiv f_1 - f_2$.

\subsection[Comparison with the first Poisson bracket $r$-matrix]{Comparison with the f\/irst Poisson bracket $\boldsymbol{r}$-matrix}\label{sec:compar}

We recall that the f\/irst Poisson bracket of rational Calogero--Moser model is expressed linearly in terms
of the Lax matrix $L$ following the formulation~(\ref{PB1}). The $r$-matrix takes the form
\begin{gather*}
r = \sum_{i \neq j} \frac{1}{q_i -q_j} e_{ij} \otimes e_{ji} +
\sum_{k \neq j} \frac{1}{q_k -q_j} e_{kk} \otimes e_{kj}.
\end{gather*}

It is interpreted~\cite{ACF2,ACF1,NAR, Sur} as a combination $r_{12} = d_{12} +c_{12}$ of two matrices realizing with
$b_{12} \equiv c_{21}$ and $a = d+c-b$ a classical semi-dynamical ref\/lection algebra corresponding to the choice
$\eps_L= 0$, $\eps_R= 1$ in (\ref{eq:dynYBE-a})--(\ref{eq:dynYBE-d}). In particular the symmetric part of $r$ is now
a sum $\frac{1}{2} (b+ c)$ of two matrices with respective weights $(0,1)$ and $(1,0)$ under adjoint action of
$\fh \oplus \fh$. It is therefore {\it not} related with the $abcd$ quadruplet realizing the second Poisson structure
of the rational CM Lax matrix, although the $d$ matrices themselves are identical. The second CM bracket is therefore
not realized as a Li--Parmentier-type quadratization of the f\/irst CM bracket. This counterexample arising in
the simplest available situation for dynamical $r$-matrices thus eliminates any possibility of extending directly
the Sklyanin--Li--Parmentier procedure
for higher Poisson brackets to the case of dynamical $r$-matrices.

Curiously enough a quadratic Poisson bracket involving the components $a$, $b$, $c$, $d$ deduced from the $r$-matrix
of the {\it first} CM bracket does exist: it arises in the formulation of the {\it first} PB of the
Ruijsenaars--Schneider rational Lax matrix~\cite{RS,Sur}.

\section{Open questions}\label{section5} 

The formulation of a simple algebraic relation \`a la Sklyanin between f\/irst and second Poisson bracket
structure in the context
of Lax matrices where dynamical $r$-matrices arise in the linear expression of their f\/irst PB seems thus, if
not altogether excluded, at least unreachable at the moment. Other issues remain open at this time:
\begin{itemize}\itemsep=0pt
  \item Def\/ine the crossed $r$-matrix formulation for $L$ and $Q$, and the $r$-matrix structure
for $Q$ when $N=2$. In this way a complete understanding
of the second PB structure including the $\{I,J\}$ and $\{J,J\}$ brackets, will be achieved.
  \item Def\/ine the $r$-matrix formulation describing the second PB structure for $N=2$ Ruijsenaars--Schneider
model (starting with the rational case). In fact a form for this second PB
structure has been conjectured~\cite{AvAnJe} but not explicitly built in terms of Lax matrix (only in terms
of trace invariants extending naturally the variables $I$ and $J$). One conjectures here that some cubic-$l$
dependent form will be relevant.
  \item Def\/ine the $r$-matrix formulation for $N=3$. The complexity of the expressions for the coordinate
Poisson brackets seems to present a dif\/f\/icult technical challenge here. This technical complexity
indicates in any case that the $r$-matrix quadruplet in a postulated quadratic form will exhibit
a dependence on both $p$ and $q$ variables, suggesting that the dynamical dependance here goes beyond the
Gervais--Neveu--Felder formulation (\ref{GNF}). A~similar issue arose some time ago~\cite{Bra} for $N \geq 4$
elliptic Calogero--Moser Lax formulation without spectral parameter, and has not been satisfactorily solved since.
The issue of $p$, $q$ dependant $r$-matrices is in any case a yet mostly unexplored one which we hope
to come back to in a near future.
\end{itemize}

\subsection*{Acknowledgements}
This work was sponsored by CNRS, Universit\'e de Cergy-Pontoise,
Universit\'e de Savoie and ANR Project
DIADEMS (Programme Blanc ANR SIMI1 2010-BLAN-0120-02). J.A.~wishes to
thank LAPTh for their kind hospitality. We thank the referees for their helpful
queries.

\pdfbookmark[1]{References}{ref}
\LastPageEnding

\end{document}